\documentclass{ws-jai}
\usepackage[flushleft]{threeparttable}
\usepackage{siunitx}
\usepackage{amsmath}
\usepackage{gensymb}
\usepackage[colorlinks=true,allcolors=blue]{hyperref}
\usepackage{graphicx}
\usepackage{caption}
\usepackage{subcaption}
\usepackage[bottom]{footmisc}
\usepackage{url}
\usepackage{natbib}

\def\albatros{ALBATROS}
\def\prizm{PRI$^{\rm Z}$M}

\begin{document}

\catchline{}{}{}{}{} 

\markboth{H.~C.~Chiang et al.}{ALBATROS instrument}

\title{The Array of Long Baseline Antennas for Taking Radio
  Observations from the Sub-Antarctic}

\author{H.~C.~Chiang$^{1,2,\dagger}$, T.~Dyson$^{1}$, E.~Egan$^{1}$,
  S.~Eyono$^{2}$, N.~Ghazi$^{2}$, J.~Hickish$^{3}$,
  J.~M.~J\'auregui--Garcia$^{4}$, V.~Manukha$^{5}$, T.~M\'enard$^{1}$,
  T.~Moso$^{2}$, J.~Peterson$^{6}$, L.~Philip$^{7}$,
  J.~L.~Sievers$^{1,7}$, S.~Tartakovsky$^{1}$}

\address{
$^{1}$Department of Physics, McGill University, Montr\'eal, Quebec H3A 2T8, Canada\\
$^{2}$School of Mathematics, Statistics, and Computer Science,
  University of KwaZulu--Natal, Durban 4000, South Africa\\
$^{3}$Department of Astronomy, University of California, Berkeley,
  California 94720, USA\\
$^{4}$Canadian Institute for Theoretical Astrophysics, Toronto, Ontario M5R 2M8, Canada \\
$^{5}$South African National Space Agency, Hermanus 7200, South Africa\\
$^{6}$Department of Physics, Carnegie Mellon University,
  Pittsburgh, Pennsylvania 15213, USA\\
$^{7}$School of Chemistry and Physics,
  University of KwaZulu--Natal, Durban 4000, South Africa
}

\maketitle

\corres{$^{\dagger}$Corresponding author: hsin.chiang@mcgill.ca}

\begin{history}
\received{(to be inserted by publisher)};
\revised{(to be inserted by publisher)};
\accepted{(to be inserted by publisher)};
\end{history}

\begin{abstract}
Measurements of redshifted 21-cm emission of neutral hydrogen at
$\lesssim30$~MHz have the potential to probe the cosmic ``dark ages,''
a period of the universe's history that remains unobserved to date.
Observations at these frequencies are exceptionally challenging
because of bright Galactic foregrounds, ionospheric contamination, and
terrestrial radio-frequency interference.  Very few sky maps exist at
$\lesssim30$~MHz, and most have modest resolution.  We introduce the
Array of Long Baseline Antennas for Taking Radio Observations from the
Sub-Antarctic (\albatros), a new experiment that aims to image
low-frequency Galactic emission with an order-of-magnitude improvement
in resolution over existing data.  The \albatros\ array will consist
of antenna stations that operate autonomously, each recording baseband
data that will be interferometrically combined offline.  The array
will be installed on Marion Island and will ultimately comprise 10
stations, with an operating frequency range of 1.2--125~MHz and
maximum baseline lengths of $\sim20$~km.  We present the
\albatros\ instrument design and discuss pathfinder observations that
were taken from Marion Island during 2018--2019.
\end{abstract}


\section{Introduction}

Measurements of redshifted \SI{21}{\cm} emission of neutral hydrogen
across a wide range of radio frequencies have the potential to
elucidate the universe's history from the cosmic ``dark ages'' up to
the formation of large-scale
structures~\citep[e.g.,][]{2012RPPh...75h6901P}.  The dark ages, which
occurred after recombination and correspond to a period when the
universe was filled with neutral hydrogen, are unexplored to date and
represent one of the final observational frontiers in cosmology.  The
number of independent modes that are contained in the matter power
spectrum in this epoch is orders of magnitude greater than the
corresponding number for cosmic microwave background (CMB)
measurements~\citep{2004PhRvL..92u1301L}.  Redshifted 21-cm
measurements of the dark ages therefore hold the potential to
significantly improve cosmological parameter constraints over those
currently derived from the CMB; however, the required observational
frequencies of $\lesssim 30$~MHz are exceptionally difficult to
access.  The primary experimental challenges include Galactic
foreground emission, ionospheric interference, and terrestrial
radio-frequency interference (RFI).

\begin{figure}
  \begin{center}
    \includegraphics[width=0.6\linewidth]{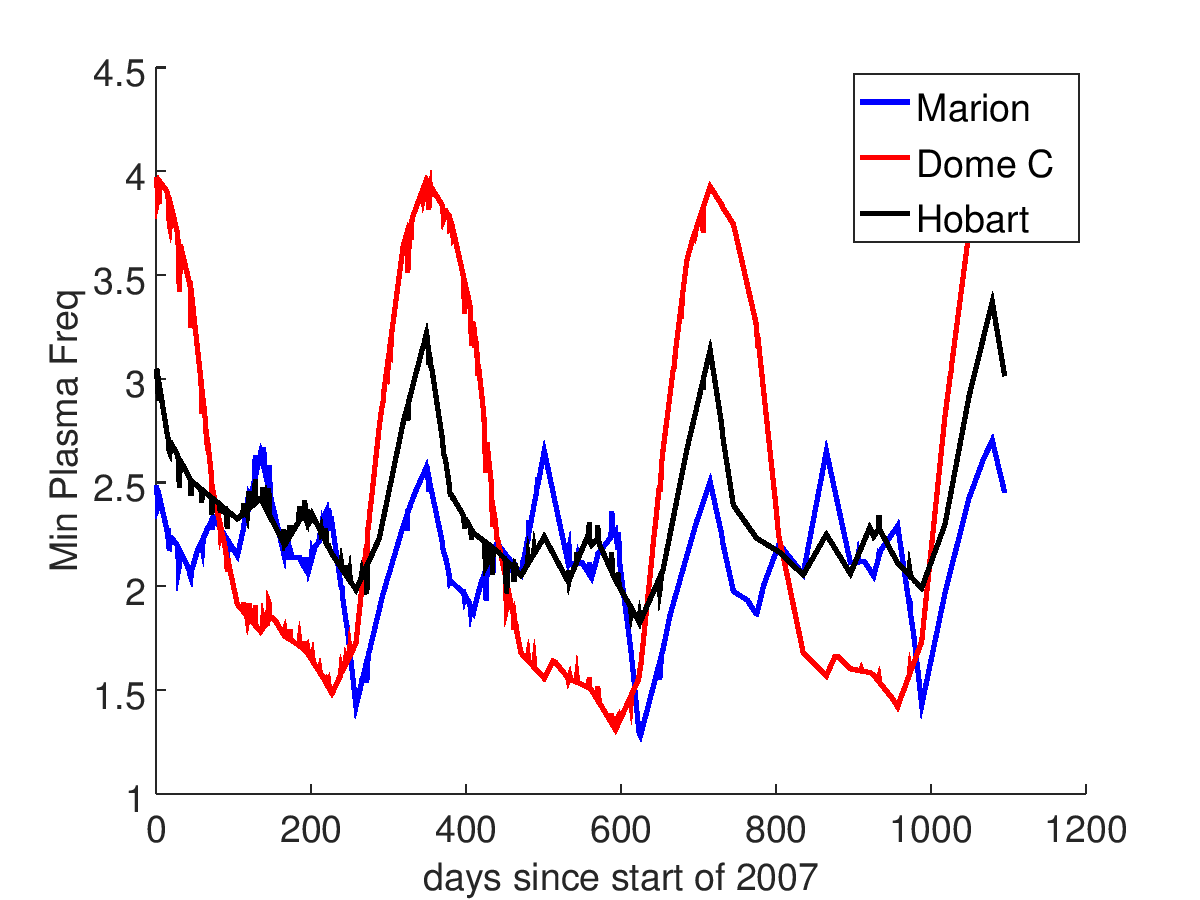}
    \caption{Minimum plasma frequency predicted by the International
      Reference Ionosphere model during the last solar minimum.  At
      some locations on Earth, the plasma frequency may drop as low as
      $\sim1.5$~MHz.}
    \label{Fig:iri}
  \end{center}
\end{figure}

Very few experiments have surveyed the radio sky at $\lesssim 30$~MHz,
and here, we highlight only a subset.  At the very lowest frequencies,
the state of the art among ground-based measurements dates from the
1950s, when Reber and Ellis caught brief glimpses of the $2.1$~MHz sky
at $\sim 5$\degree\ resolution, using an array of
192~dipoles~\citep{1956JGR....61....1R}.  A few space-based missions
have also performed measurements at similarly low frequency ranges;
for example, the Radio Astronomy Explorer-2 operated at
\SIrange{0.025}{13}{\MHz} with $\sim 10$\degree\ resolution at
\SI{4.7}{\MHz}~\citep{1975A&A....40..365A}.  Subsequent ground-based
experiments include the Dominion Radio Astrophysical Observatory
(DRAO) surveys of the northern sky at 10~MHz and 22~MHz, with
$\sim2$\degree\ and 1.1--1.7\degree\ resolution,
respectively~\citep{1976MNRAS.177..601C, 1999A&AS..137....7R}.
Low-frequency imaging experiments that are currently operational or
under construction include the Ukranian T-shaped Radio
telescope~\citep{2016JAI.....541010Z}, with 0.5\degree\ resolution at
the center of its 8--32~MHz operating range, and the New extension in
Nan\c{c}ay upgrading LOFAR~\citep{7136773}, which operates at
10--85~MHz with 1.5\degree\ resolution in standalone mode, and
improved resolution at $>30$~MHz when combined with LOFAR.  Existing
measurements with finer resolution are mainly at higher frequencies;
for example, the Owens Valley Long Wavelength Array mapped the sky
with \SI{15}{\arcminute} resolution between 36.528~MHz and
73.152~MHz~\citep{2018AJ....156...32E}.  The dearth of information
about the $\lesssim 30$~MHz sky, particularly in the southern
hemisphere, is illustrated by the Global Sky
Model~\citep{2017MNRAS.464.3486Z}, which incorporates the DRAO 10~MHz
and 22~MHz data as the only two inputs at these low frequencies with
sufficiently meaningful sky coverage and resolution.

The first step toward laying the groundwork for
possible measurements of the dark ages is obtaining an improved map of
Galactic foregrounds at $\lesssim 30$~MHz.  In addition to serving as
a stepping stone for future cosmological constraints from the dark
ages, maps at these frequencies can also shed new light on Galactic
astrophysics.  Above $\sim30$~MHz, synchrotron emission from the
Galaxy follows a power law with a frequency dependence of $\sim
\nu^{-0.7}$.  At lower frequencies, synchrotron self-absorption
becomes non-negligible, and the spectrum transitions to $\nu^{+5/2}$
dependence.  Since self-absorption causes the optical depth along the
line of sight to depend on frequency, observations at $\lesssim
30$~MHz have the potential to probe the three-dimensional cosmic ray
structure of the Galaxy~\citep{2002ApJ...575..217P}.  Low-frequency
observations can also be used to image radio recombination lines
(RRLs), which provide a means for studying cool, low-density regions
of the interstellar medium (ISM).  These absorption lines arise from
Rydberg atoms and are highly sensitive to the surrounding environment.
The RRL spectrum and absorption profiles can therefore constrain the
detailed properties of the ISM regions in which the Rydberg atoms are
formed~\citep{2009NewAR..53..259G, 2007MNRAS.374..852S}.  Finally,
observations at $\lesssim30$~MHz may provide new views of Earth's
ionosphere, which absorbs and refracts at radio frequencies and
becomes completely opaque below the plasma cutoff frequency.  The
cutoff frequency and the levels of absorption and refraction are
time-varying and spatially dependent.  Experiments that image the
radio sky at multiple frequencies bracketing the plasma cutoff can
therefore simultaneously image the ionosphere and provide spectral
riometry data for space weather studies~\citep{2014GeoRL..41.5370K}.

There are proposed efforts to perform new low-frequency measurements
from space, where there is no contamination from the ionosphere, and
the lunar farside can potentially block RFI from the
Earth~\citep{2019arXiv190710853C, 2019arXiv190804296K}.  Although the
combination of ionosphere and RFI significantly impedes low-frequency
radio observations from most locations on Earth, preliminary
observations from Marion Island~\citep{2019JAI.....850004P} suggest
that such observations may still be accessible from carefully selected
locations and with new technology developments.  In this paper, we
present the Array of Long Baseline Antennas for Taking Radio
Observations from the Sub-Antarctic (\albatros), a new experimental
effort that aims to map the low-frequency sky using an array of
autonomous antenna stations.  We describe the overall instrument
design and preliminary measurements from engineering runs that were
performed on Marion Island during 2018--2019.

\begin{figure}
    \centering
    \begin{subfigure}[t]{0.6\textwidth}
        \centering
        \includegraphics[width=\linewidth]{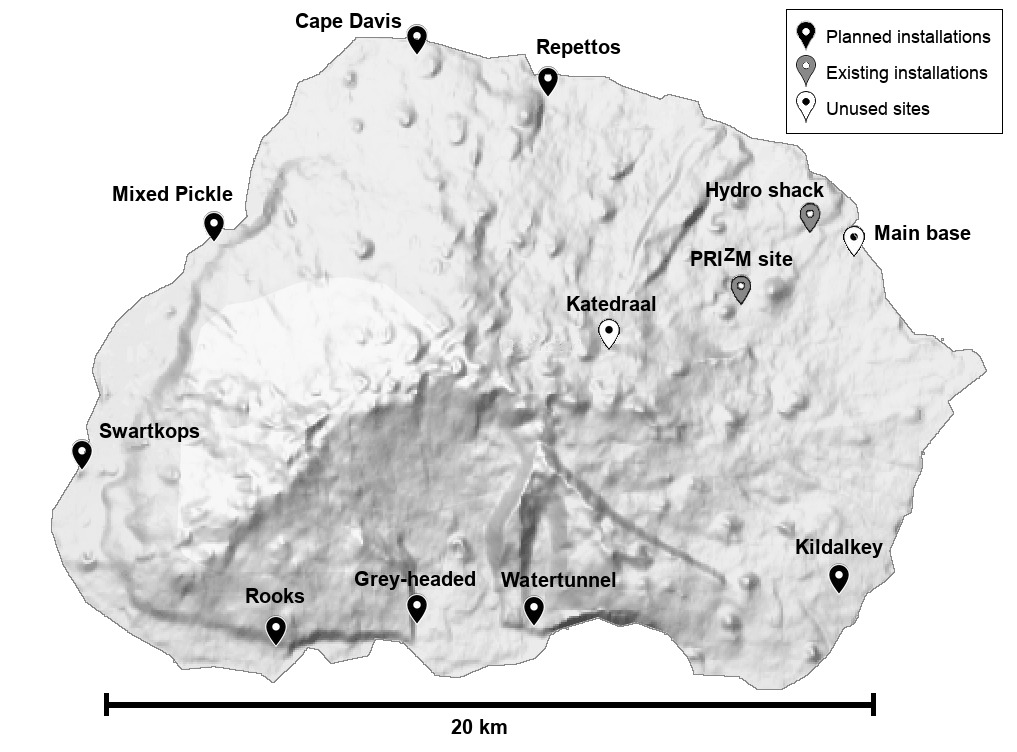} 
        \caption{} \label{Fig:marion_map}
    \end{subfigure}
    \hfill
    \begin{subfigure}[t]{0.39\textwidth}
      \centering
        \includegraphics[width=\linewidth]{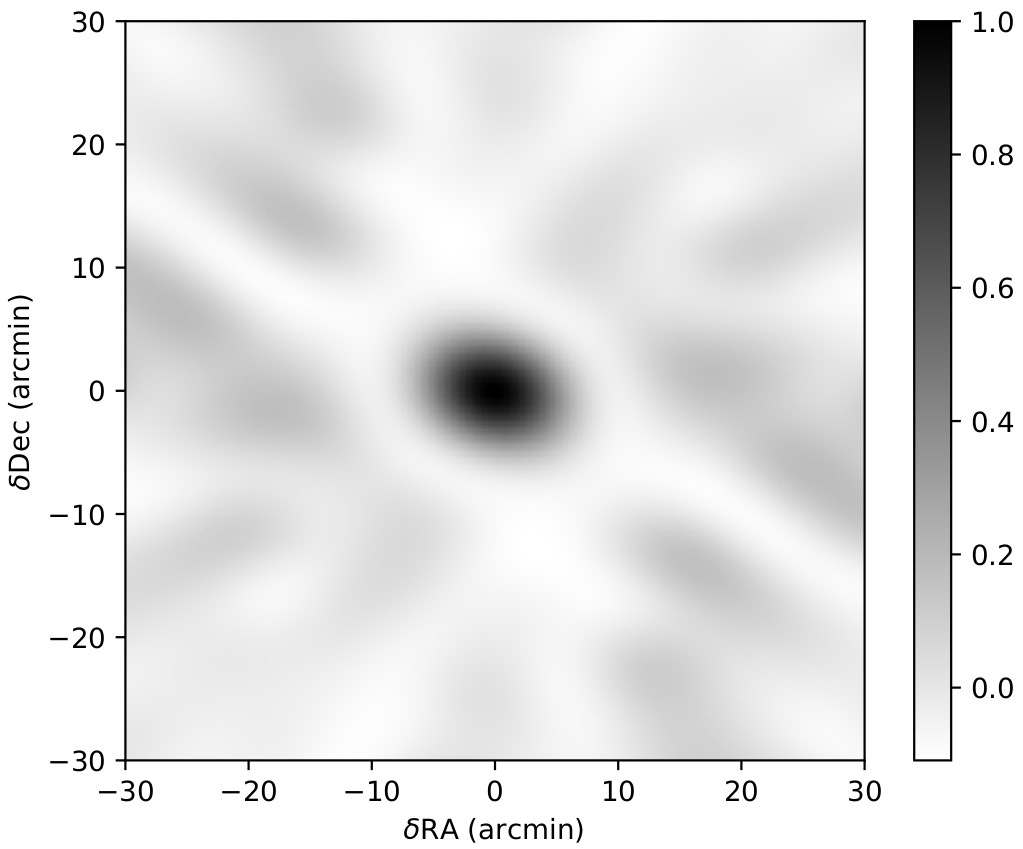}
        \caption{} \label{Fig:marion_beam}
    \end{subfigure}
    \caption{{\bf (a)} Map of Marion Island.  The
      \albatros\ pathfinder antennas are currently installed at the
      \prizm\ site and at the hydro shack (gray markers).  The black
      markers denote the eight coastal huts, which will be used for
      future \albatros\ antenna installations.  The white markers
      denote other available infrastructure points that will not be
      used for antennas.  {\bf (b)} Synthesized beam at 5~MHz at zenith from the
      full \albatros\ array, using the existing and planned
      installation locations shown on the map.  Using an octave of
      bandwidth spanning 3.5--7~MHz in a single snapshot, we obtain a synthesized beam with 
      a full width of $\sim7'$.}\label{Fig:marion_map_beam}
\end{figure}

\section{ALBATROS overview}

The primary requirements that drive the design for a low-frequency
imaging experiment are 1) desired resolution, 2) low RFI, and 3) quiet
ionospheric conditions.  As a benchmark, an interferometer operating
at 30~MHz requires baseline lengths of $\sim0.5$--1~km 
to achieve a resolution of $\sim 1$\degree.  This length scales
inversely with frequency, and therefore $\sim 10$-km lengths are
required at few-MHz frequencies to improve substantially upon the resolutions
achieved to date.  The experiment installation site must be remote to
keep RFI to a minimum, and polar or near-polar latitudes generally
have lower ionospheric plasma frequency cutoffs relative to other
locations on Earth.  \autoref{Fig:iri} illustrates predictions from
the International Reference Ionosphere
model~\citep{2018AdRS...16....1B} for the minimum plasma frequency
during the last solar minimum, which started in approximately 2007.
Three locations are shown: Marion Island, the focus of this work;
Dome~C in Antarctica, which is another isolated location used for
astronomical observations; and Hobart, Tasmania, where Reber performed
his $\sim2$~MHz observations.  The model predictions illustrate that
the ionospheric plasma cutoff frequency at Marion Island may drop as
low as $\sim1.5$~MHz during solar minima.  Since we are currently
experiencing another solar minimum~\citep{2018NatCo...9.5209B}, the
timing is opportune for new low-frequency observations.

Marion Island is a research base that is located in the southern
Indian Ocean at \ang{46;54;45}S, \ang{37;44;37}E and is operated by
the South African National Antarctic Programme.  The island lies
roughly \SI{2000}{\kilo\metre} from the nearest continental landmasses
and has an exceptionally quiet RFI
environment~\citep{2019JAI.....850004P}.  As illustrated in
\autoref{Fig:marion_map}, Marion has an area of 335~km$^2$ and can
therefore support antenna installations with $>10$-km baseline
lengths.  The main Marion base is located on the northeast side of the
island, and there are eight rest huts along the coast (Cape Davis,
Grey-headed, Kildalkey, Mixed Pickle, Repettos, Rooks, Swartkops,
Watertunnel) and one in the interior (Katedraal) that can serve as
existing infrastructure points for antenna installations.  The planned
\albatros\ installation sites include the coastal huts, but exclude
the main base and Katedraal for RFI and accessibility reasons,
respectively.  \autoref{Fig:marion_map} also shows the locations of
the \albatros\ pathfinder antennas that are currently installed at the
\prizm\ site and at the hydro shack.

Using the eight coastal huts, the \prizm\ site, and the hydro shack as
the nominal \albatros\ installation locations, the computed
synthesized beam is shown in \autoref{Fig:marion_beam}.  The beam
width at 5~MHz is roughly \SI{7}{\arcminute}, which represents over an
order of magnitude improvement in resolution over other existing
measurements.  One of the challenges in constructing an interferometer
array on Marion Island is that the rugged terrain precludes the
possibility of directly cabling and correlating antennas across large
distances.  The final \albatros\ antenna stations will therefore
operate {\it autonomously}, recording baseband data over extended
periods of time for subsequent offline correlation.  We have conducted
two engineering runs: 1) a two-element, directly correlated pathfinder
to qualitatively understand the sky signal, and 2) a single station to
test the readout and power handling technology that are required for
autonomous operation.

\begin{figure}
  \begin{center}
    \includegraphics[width=0.7\linewidth]{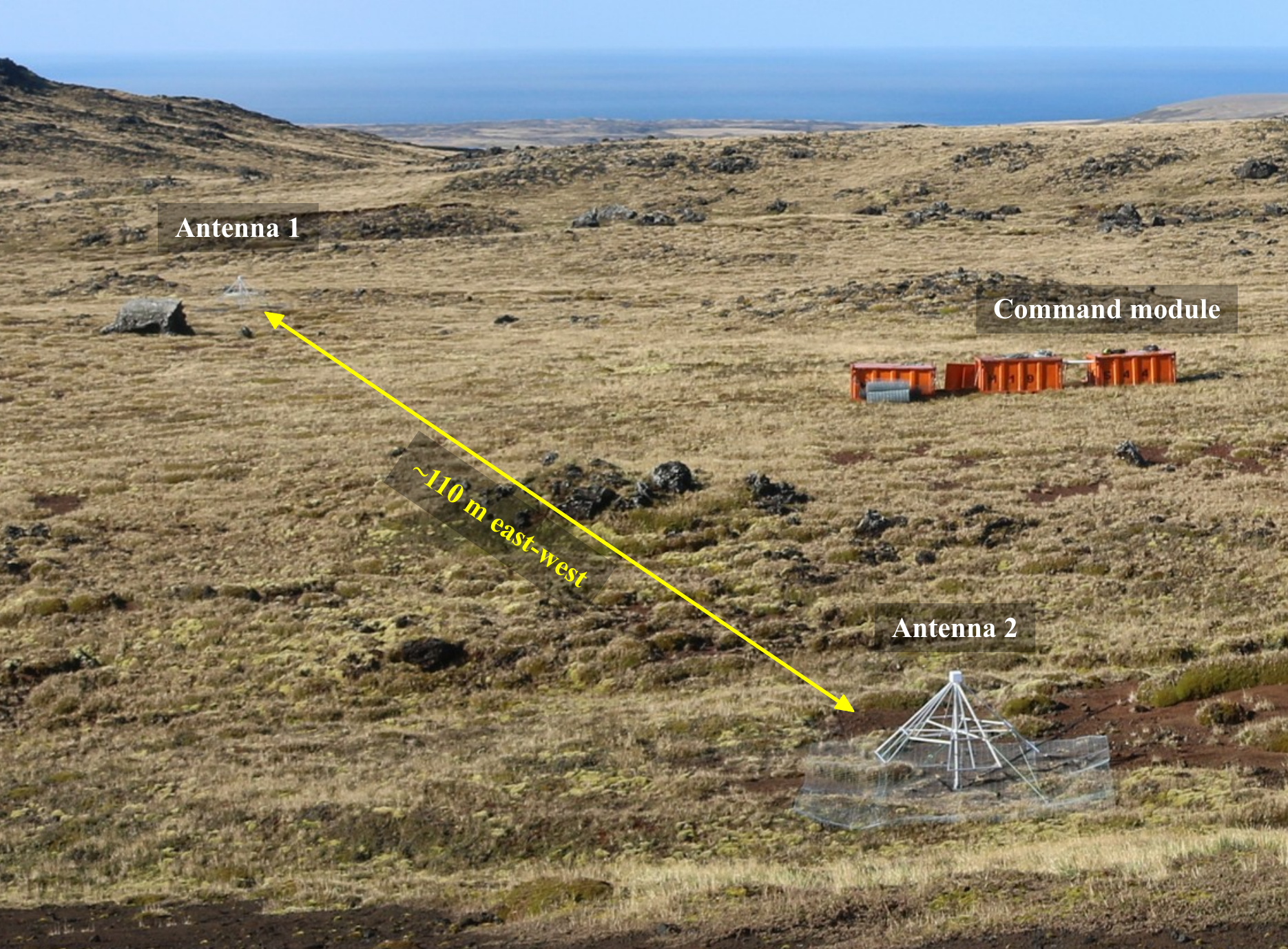}
    \caption{The two-element, directly correlated
      \albatros\ pathfinder installed at the \prizm\ site.  Two
      dual-polarization antennas are separated by roughly 110~m on an
      east--west baseline. Coaxial cables connect the antennas to a
      shipping container that houses the readout electronics and
      serves as the ``command module.''}
    \label{Fig:albatros2}
  \end{center}
\end{figure}

\section{Two-element pathfinder}\label{s:2elem}

\begin{figure}
  \begin{center} \includegraphics[width=1.0\linewidth]{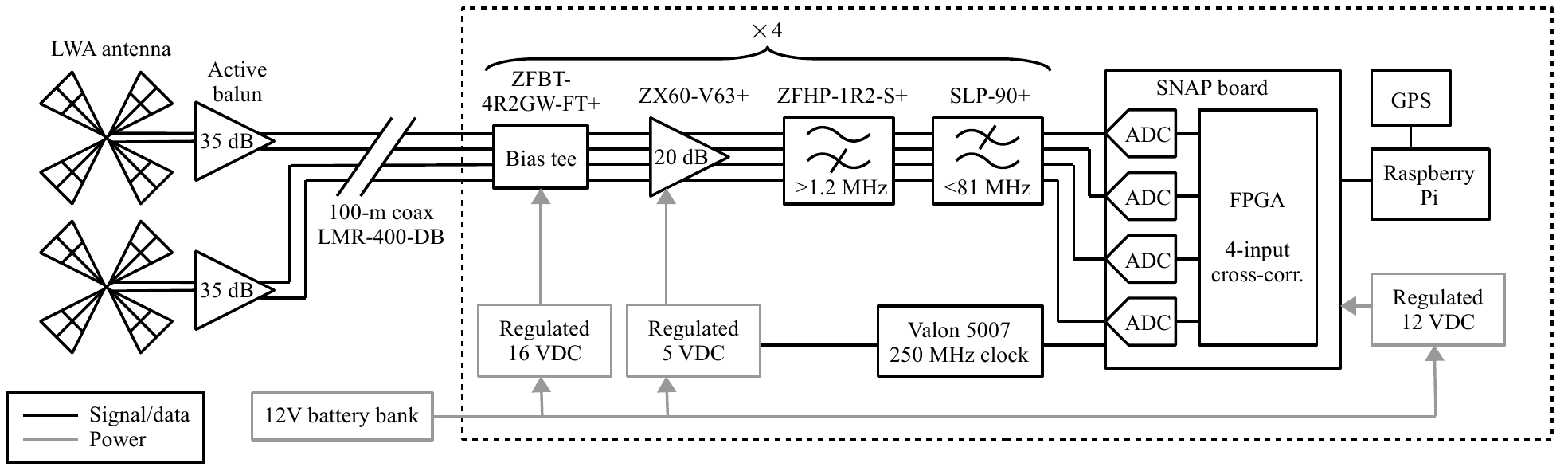}
    \caption{Two-element \albatros\ pathfinder block diagram.  Signals
      from two dual-polarization LWA antennas are amplified by
      front-end active baluns~\citep{2012PASP..124.1090H}.  The
      antennas are connected via 100-m coaxial cables to the back-end
      readout electronics, which are housed in a Faraday cage denoted
      by the dashed box.  Each of the four antenna outputs is passed
      to a second-stage electronics chain consisting of filters and
      further amplification.  The signals are digitized at 250~Msamp/s
      by a SNAP board, which includes an on-board FPGA that computes
      auto- and cross-spectra from and between the four inputs.  A
      Raspberry Pi controls the SNAP board and saves the data.}
    \label{Fig:albatros2_schem}
  \end{center}
\end{figure}

The first exploratory \albatros\ measurements were conducted with a
two-element pathfinder that used direct correlation (without
autonomous operation).  \autoref{Fig:albatros2} illustrates this
pathfinder, which was installed at the \prizm\ site (\ang{46;53;13}S,
\ang{37;49;10.7}E) in April 2018.  The system block diagram is shown
in \autoref{Fig:albatros2_schem}, and each of the subsystems is
described in detail below.

\subsection{Antenna}\label{s:antenna}

We employ two dual-polarization Long Wavelength Array (LWA) dipole
antennas~\citep{Memo28}.  The LWA antennas have a long development
history, are well characterized, and are simple to install and
physically robust.  The antennas form an east--west baseline with a
separation of \SI{110}{m}, and the polarizations are aligned with the
cardinal directions.  Welded wire mesh screens, roughly 3~m on a side,
are installed on the ground below the antennas.  The beam pattern is
roughly omnidirectional with a solid angle that varies between 2.2 and
2.7 steradians over 5--100~MHz.  Although the LWA antenna is not
optimized for observations below 10~MHz, using this standard antenna
allowed an initial test of the site and of the interferometer systems.
For the final \albatros\ array, we are exploring alternative antenna
designs that will cover the full frequency range.

\subsection{Front-end active balun}\label{s:fee}

The LWA active-balun front-end electronics (FEE) circuit uses a
Mini-Circuits GALI-74+ MMIC to amplify each dipole leg against ground,
presenting each leg with a 50$\ohm$ impedance and providing a nominal
gain of 25~dB. The two GALI-74+ outputs are differenced using a
passive {180\degree} hybrid coupler. The coupler output is filtered by
a $\sim$150~MHz lowpass and receives an additional 12~dB of gain from
a Mini-Circuits GALI-6+ MMIC. This last amplifier drives the output
signal onto a 100-m 50$\ohm$ coaxial cable. Thus, neglecting mismatch
loss between the dipoles (electrically small in the frequency range of
interest) and the 100{\ohm} input impedance, as well as the hybrid
insertion loss ($<$1dB), the front-end electronics provide a gain of
$\sim$37~dB~\citep{2012PASP..124.1090H}.  Each FEE circuit is powered
by 16~V, which is fed on the coaxial cable through a bias tee.  The
lengths of the coaxial cables are the same for both antennas, thus
placing the phase center at zenith.

\subsection{Back-end electronics}

The back-end readout electronics are housed in a Faraday cage located
100~m away from the antennas to mitigate possible self-generated RFI.
Each of the four antenna signals is passed to a second-stage
electronics chain consisting of an amplifier (Mini-Circuits ZX60-V63+),
and a pair of high- and low-pass filters (Mini-Circuits ZFHP-1R2+ and
SLP-90+) that together band-limit the signal to 1.2--\SI{81}{MHz}.
The amplifier has a nominal gain of 20~dB, and the high- and low-pass
filters contribute nominal insertion losses of 0.2~dB and 0.14~dB,
respectively.

A Smart Network ADC Processor~\citep[SNAP;][]{2016JAI.....541001H}
board digitizes the RF signals at 250~Msamp/s and uses a Xilinx
Kintex~7\footnote{\url{http://www.xilinx.com/products/silicon-devices/fpga/kintex-7.html}}
FPGA to calculate full cross-correlations of the four inputs,
producing four auto- and six cross-spectra as outputs, over 2048
channels spanning the frequency range 0--125~MHz.  The channelization
is performed with a
CASPER\footnote{\url{https://casper.berkeley.edu/}}-based polyphase
filter bank, and the spectra are accumulated over $\sim4$-s intervals.
A Valon 5007 frequency synthesizer module provides the clock signal
for the SNAP board.  A Raspberry Pi (RPi)~3B+ single board computer
controls the SNAP board and receives the spectra via GPIO connections.
The data rate of the averaged spectra is roughly 400~MB/day (with
compression), and this low volume allows the spectra to be saved to
the RPi on-board SD card.  An Adafruit Ultimate GPS
module\footnote{\url{https://www.adafruit.com/product/746}}, connected
to an active external GPS antenna, provides absolute timing for the
RPi.

\subsection{Power}

A bank of four 12-V, 200-Ah AGM batteries, connected in parallel,
powers the two-element pathfinder system.  The batteries are manually
charged with a Honda EU30is generator that is housed on site.  The
total power draw of the two-element pathfinder is $\sim45$~W, and when
fully charged, the battery bank can power the system for about a week.
The raw battery voltage is passed to several DC/DC converters that
supply regulated voltages to the SNAP board, FEE, amplifiers and the
clock.

\section{Single autonomous station pathfinder}\label{s:autonomous}

\begin{figure}
    \centering
    \begin{subfigure}[t]{0.48\textwidth}
        \centering
        \includegraphics[width=\linewidth]{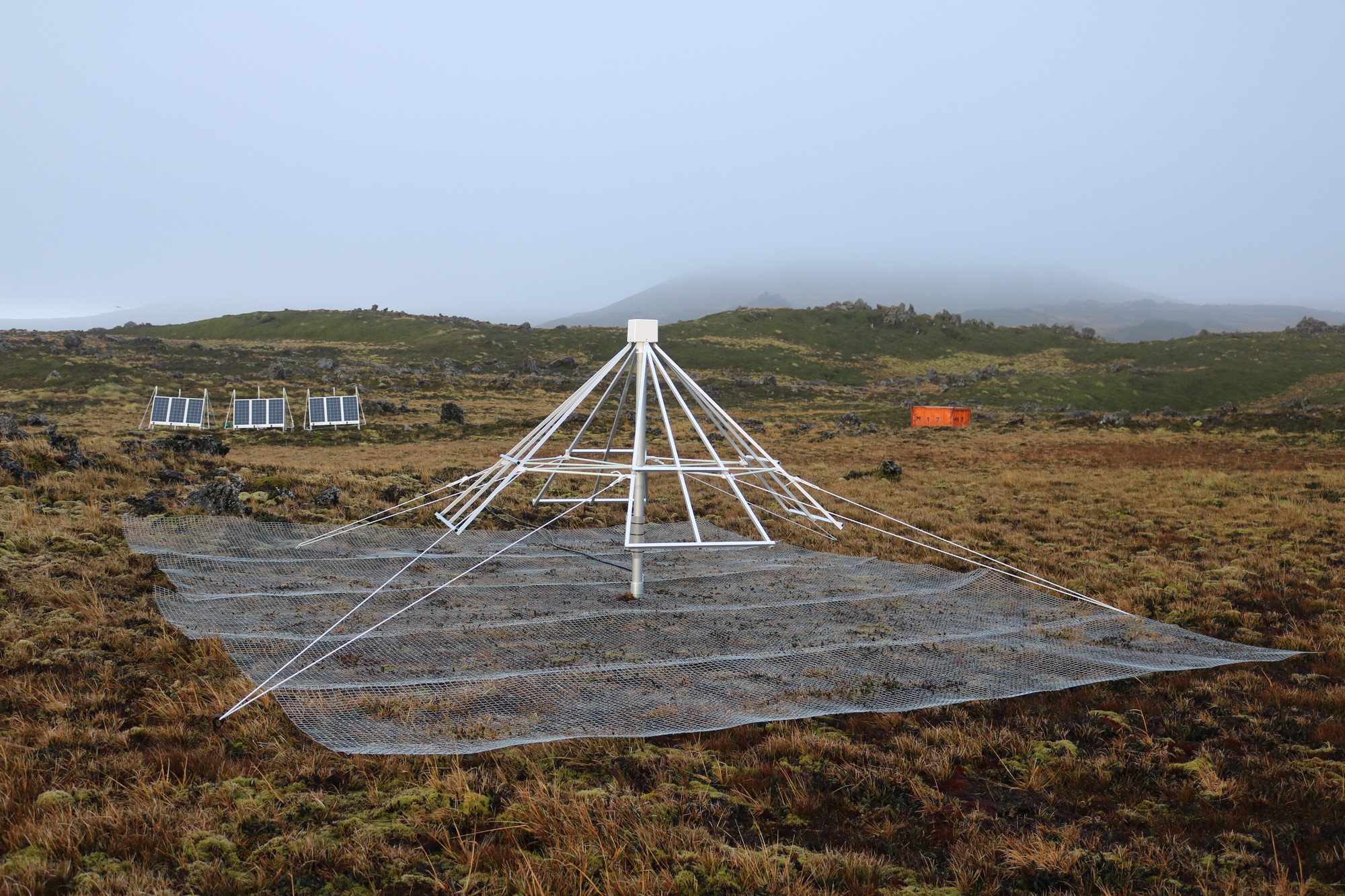} 
        \caption{} \label{Fig:autonomous_antenna}
    \end{subfigure}
    \hfill
    \begin{subfigure}[t]{0.48\textwidth}
      \centering
        \includegraphics[width=\linewidth]{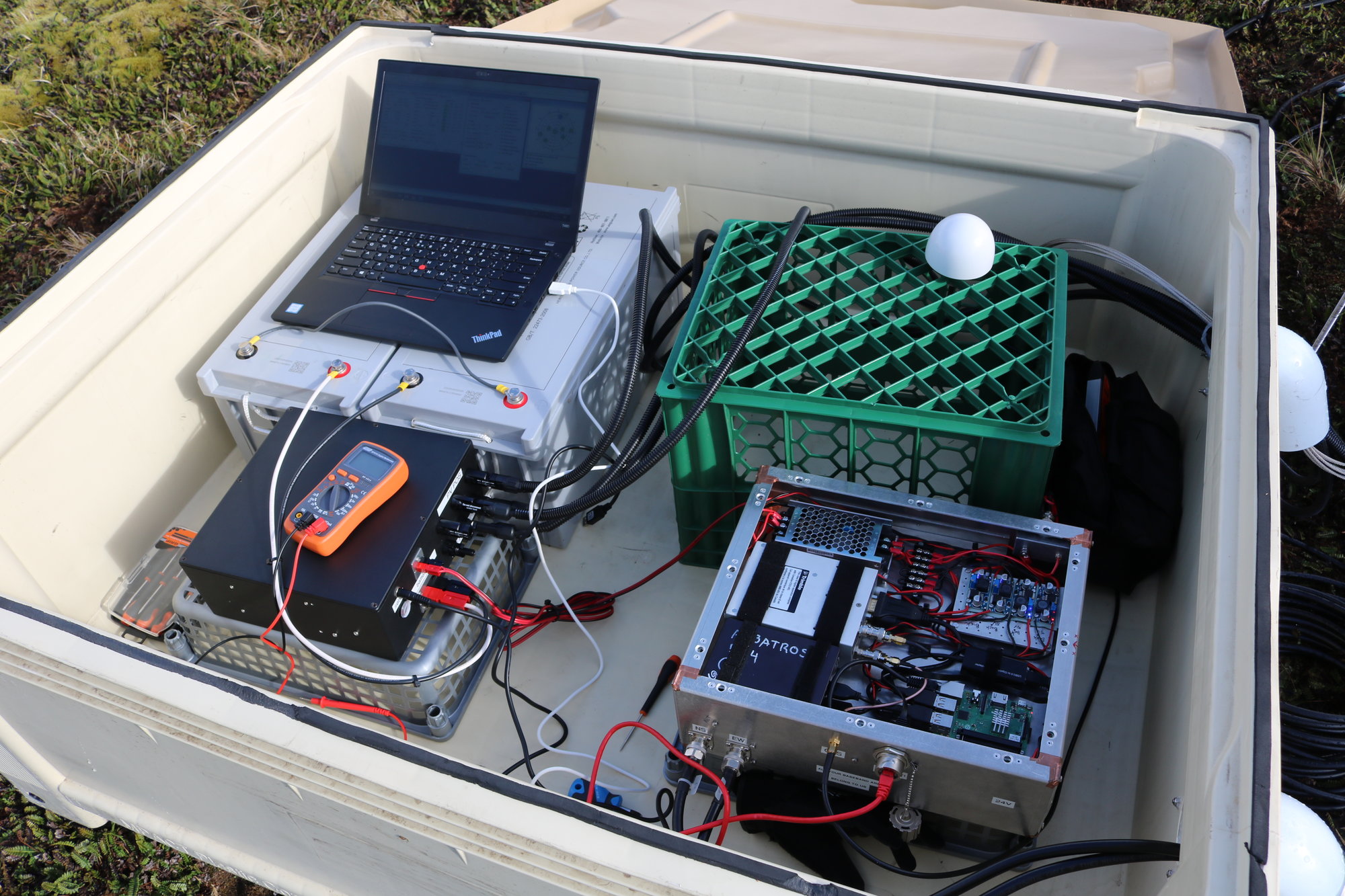}
        \caption{} \label{Fig:autonomous_electronics}
    \end{subfigure}
    \caption{{\bf (a)} Single autonomous station pathfinder installed
      at the hydro shack site.  The system is powered by a bank of
      solar panels that are visible in the background. {\bf (b)} A
      weather-proof container, which sits near the solar panels,
      houses the batteries, solar charge controller, and readout
      electronics.}\label{Fig:autonomous}
\end{figure}

\begin{figure}
  \begin{center}
    \includegraphics[width=\linewidth]{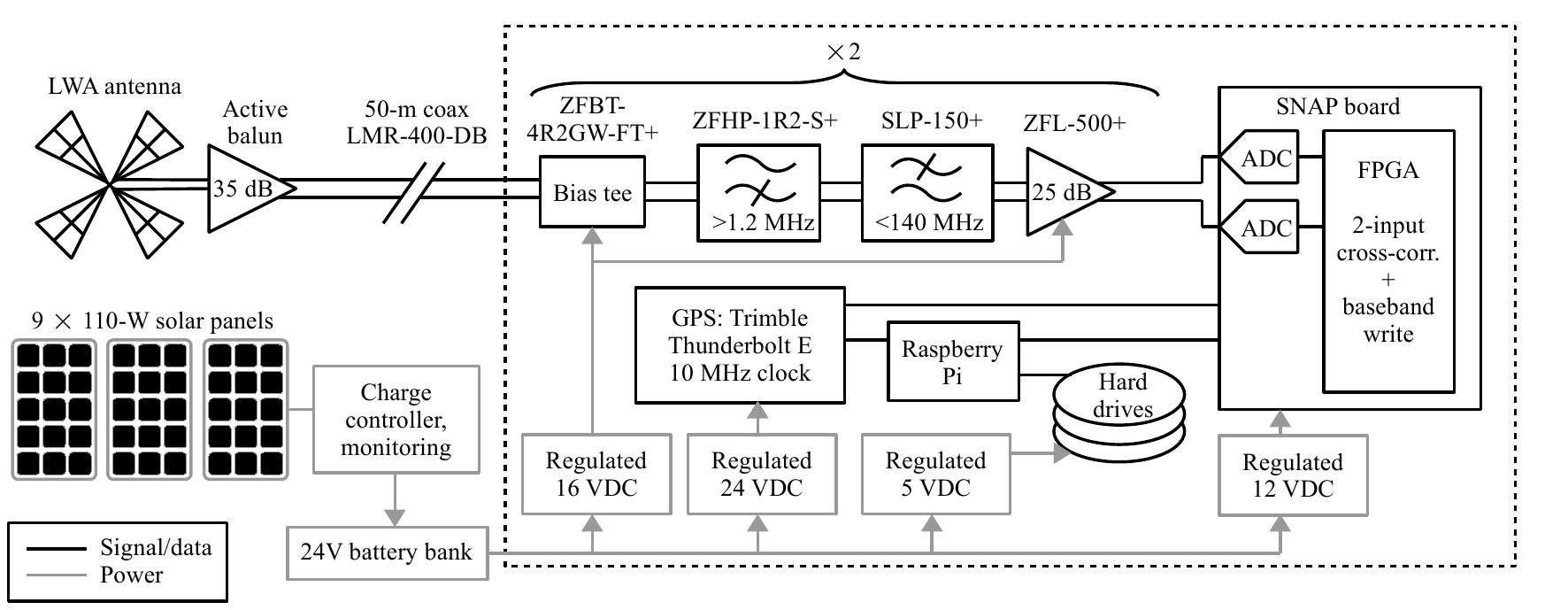}
    \caption{Single-antenna autonomous station block diagram.  A
      dual-polarization LWA antenna, equipped with a front-end active
      balun, connects via 50-m coaxial cables to the back-end readout
      electronics, housed in a Faraday cage denoted by the dashed box.
      Each of the two antenna signals is passed to a second-stage
      electronics chain consisting of filters and further
      amplification.  The signals are digitized at 250~Msamp/s by a
      SNAP board, which includes an on-board FPGA that computes
      channelized baseband and spectra from both inputs.  A Raspberry
      Pi controls the SNAP board and receives the baseband data and
      spectra.  The baseband data are saved to external hard drives.
      The system is powered by solar panels that charge a 24-V battery
      bank.}
    \label{Fig:albatros1_schem}
  \end{center}
\end{figure}

The full \albatros\ configuration will comprise an array of 10
autonomous antenna stations, each recording baseband over tunable
frequency windows (with 10--20~MHz total bandwidth) within the full
0--125~MHz operating range.  Baseband data will be periodically
physically retrieved from each station and subsequently correlated
offline.  The \albatros\ stations, located at the eight coastal hut
sites plus the hydro shack and \prizm\ site, will be separated by
baselines of $\sim20$~km as shown in
\autoref{Fig:marion_map_beam}. One fully autonomous
\albatros\ station, shown in \autoref{Fig:autonomous}, was deployed in
April 2019 at the hydro shack location (\ang{46;52.205;}S,
\ang{37;50.612;}E) on Marion Island as a first step in testing the
technology needed to establish the full array.  The system block
diagram is shown in \autoref{Fig:albatros1_schem}.  The antenna and
front-end active balun in the single autonomous station are identical
to those used in the two-element pathfinder (\S\ref{s:antenna} and
\S\ref{s:fee}), and the back-end electronics and power system are
described in detail below.

\subsection{Back-end electronics}\label{s:autonomous_backend}

The back-end readout electronics are housed in a Faraday cage located
50~m away from the antennas. The analog signal chain consists of a
pair of high- and low-pass filters (Mini-Circuits ZFHP-1R2+ and
SLP-150+) that together band-limit the signal to 1.2--140~MHz, and the
filters are followed by a Mini-Circuits ZFL-500+ amplifier.  In
contrast to the two-element pathfinder, the low-pass cutoff is
increased from 81~MHz to 140~MHz to capture downlink signals at
137--138~MHz from the ORBCOMM satellite constellation.  For the final
\albatros\ array, the ORBCOMM signals provide a convenient means for
synchronization across the antenna stations, serving as a backup to
the GPS timing discussed below.  Preliminary lab tests suggest that on
time scales of $\sim30$~s, relative timing between different antenna
stations can be measured to an accuracy of better than a few tenths of
a nanosecond using a single satellite.  The corresponding phase error
at 10~MHz is $\lesssim1^\circ$.  With publicly available orbits and
multiple satellites typically within the field of view, we expect that
ORBCOMM baseband data saved at the same time as the astronomical data
can provide offline synchronization of the \albatros\ stations.
Because the ORBCOMM and science data are recorded at the same time by
the same system, improvements to the timing calibration can be applied
in post-processing.

As with the two-element pathfinder, a SNAP board digitizes each of the
two RF signals at 250~Msamp/s. In the autonomous station
configuration, the SNAP board ADCs are locked to a 10-MHz reference
produced by a Trimble Thunderbolt~E GPS-disciplined clock module.  The
SNAP board FPGA computes two data products: 1) channelized baseband
data for each polarization over tunable frequency windows within the
0--125~MHz operating range, with the options of 1-, 2-, or 4-bit
compression, and 2) auto- and cross-spectra from the two polarizations
over the full 0--125~MHz span, accumulated over few-second intervals.
The reduction in bit depth happens only \textit{after} the SNAP board
has channelized the baseband, and the cross-channel leakage (due to,
e.g., RFI and slopes in RF power as seen by the ADCs) is therefore
unaffected by the low bit depth.  An RPi~3B+ controls the SNAP board
and and receives the auto- and cross-spectra via GPIO connections, and
the spectra are saved to an on-board SD card.  The baseband data are
passed from the SNAP board to the RPi via ethernet and written to
external hard drives.  The introduction of gigabit ethernet with the
RPi~3B+ model has enabled the high data throughput associated with
writing baseband.  As a benchmark, 1-bit baseband recording of two
polarizations over 10~MHz of bandwidth yields an approximate data rate
of 5~MB/s, or 0.4~TB/day.

\subsection{Correlation}

\begin{figure}
  \begin{center}
    \includegraphics[width=0.75\linewidth]{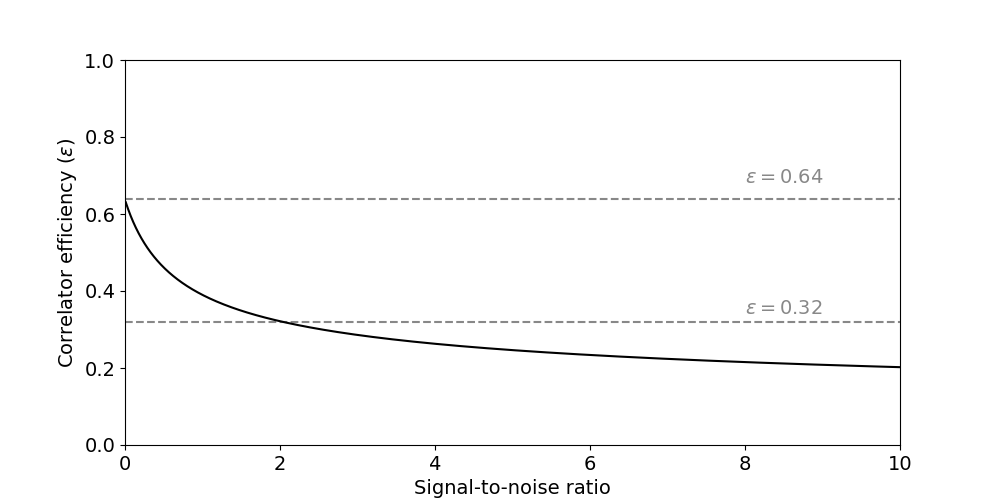}
    \caption{Correlator efficiency $\epsilon$ for a 1-bit correlator
      as a function of signal-to-noise ratio (SNR).  The efficiency is
      roughly 0.64 at low SNR and drops monotonically as the SNR
      increases.  The decrease in $\epsilon$ is relatively gentle,
      changing by only a factor of 2 for SNR $\sim2$.}
    \label{fig:1bit_efficiency}
    \end{center}
\end{figure}

The autonomous station can be configured to record 1-, 2-, or 4-bit
baseband.  Signal fidelity improves with higher numbers of bits, but
the penalty is increased data volume.  Here we discuss the prospects
of employing 1-bit correlation for the final \albatros\ array to keep
the baseband data volume to a minimum.  While the basics of 1-bit
correlation are well known in the limit where the signal level is much
lower than the noise, \albatros\ may be in a regime where the signal
level is non-negligible.  The fundamental output of a 1-bit correlator
for real data is
\begin{equation}
  x_{ij} \equiv \left < \tilde{E}_i \tilde{E}_j \right >.
\end{equation}
Here $\tilde{E}_i$ is the quantized version of the underlying electric
field $E_i$ for antenna index $i$:
\begin{equation}
  \tilde{E}_i=
  \begin{cases}
    1 & \text{if}\ E_i > 0 \\
    -1 & \text{if}\ E_i < 0.
  \end{cases}
\end{equation}
Complex data can be handled as the combination of real components.
The output is nonlinear in the underlying true signal and noise
levels; in the limit of perfectly correlated electric fields, $x_{ij}$
saturates at unity.  Recovering the true sky signals requires
inverting this nonlinear relation using the Van Vleck
corrections~\citep{1446497}.  For a 2-level correlator, the
relationship between $x_{ij}$ and the sky signals
is~\citep{1989ASPC....6...59D}
\begin{eqnarray}
\label{eqn:1bit_output}
\left < \sin\left(\frac{\pi}{2}x_{ij}\right)\right > = \frac{V_{ij}}{\sqrt{V_{ij}+N_i}\sqrt{V_{ij}+N_j}},
\end{eqnarray}
where $V_{ij}$ is the true sky visibility, and $V_{ij}+N_{i}$ and
$V_{ij}+N_{j}$ are the total, signal-plus-noise, power levels measured
by antennas $i$ and $j$, respectively.  Inverting this relation to
obtain $V_{ij}$ requires knowing the total power levels, which cannot
be determined from 1-bit data alone.  However, as described in
\S\ref{s:autonomous_backend}, the SNAP board calculates and records
auto-spectrum data products, which provide power level information on
few-second time scales.  Changes in the power levels occur on
significantly longer time scales, so the true sky visibilities can be
derived by combining cross-correlation data between different antenna
stations and the auto-spectrum data from each individual station.

To assess the loss of sensitivity as the signal level increases, we
use \autoref{eqn:1bit_output} to derive the correlator efficiency
$\epsilon$, which is the ratio of measured signal-to-noise ratio (SNR)
to the ideal, infinite-precision SNR.  \autoref{fig:1bit_efficiency}
illustrates that the decrease in $\epsilon$ as a function of SNR is
relatively shallow.  When the SNR is near zero, $\epsilon$ approaches
$2/\pi \approx 0.64$, and if the signal level increases to two times
the noise level, $\epsilon$ decreases by only a factor of $\sim2$.
The final \albatros\ array will have baselines that are long relative
to the observing wavelength, and the correlated component of the
electric field (which sets the relevant signal level) is likely a
small fraction of the total received power.  This qualitative argument
can by made by considering antennas in the $uv$~plane in the flat-sky
approximation.  Any individual antenna receives all power present in
the $uv$~plane (plus system noise), but the correlated contribution
comes only from the portion of the $uv$~plane that is encompassed by
the primary beam at the antenna's $uv$-space coordinate.  Assuming
that the sky can be described as a Gaussian random field, the maximum
correlated signal between two antennas is determined by the aperture
filling factor of that pair.  For the case of dipole antennas, the
filling factor is the square of the wavelength divided by the baseline
length.  Therefore, in the long-baseline limit, the correlated signal
is small in comparison to the total power received, and the correlator
efficiency is unlikely to decrease significantly from the ideal
$2/\pi$ value.  This efficiency behavior remains true even when the
brightest radio sources are visible, since the primary beam
encompasses more than a steradian solid angle (the sun is irrelevant
since ionospheric effects restrict \albatros\ to night-time
observing).  Centaurus~A~\citep{culgoora1} and M1~\citep{Braude1969}
affect the total power expected from the Global Sky
Model~\citep{gsm2008} at only the $\sim$10\% level.  Additionally, as
will be shown in \S\ref{s:obs}, preliminary \albatros\ data
demonstrate that the visibilities are only a few percent of the total
power amplitude, thus indicating that diffuse emission dominates even
the brightest sources over the large primary beam.

\begin{figure}[t]
  \begin{center}
    \includegraphics[width=\linewidth]{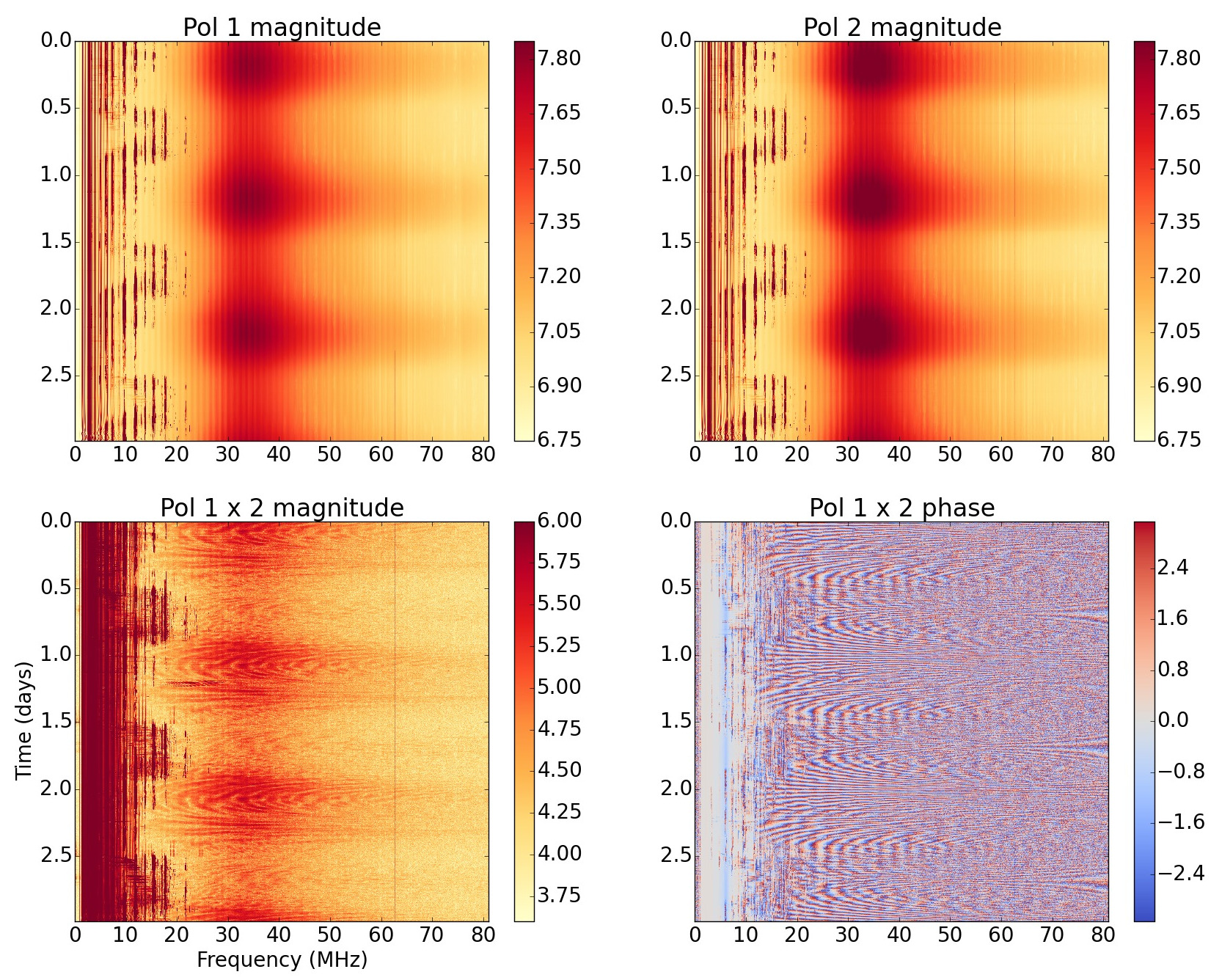}
    \caption{Spectra from a co-aligned polarization pair from the
      two-element \albatros\ pathfinder, shown as a function of
      frequency and time.  The auto- and cross-spectra are shown in
      the top and bottom rows, respectively.  The spectrum magnitudes
      are in uncalibrated ADC units on a logarithmic scale, and the
      cross-spectrum amplitude is about two orders of magnitude
      fainter than the autospectra.  The cross-spectrum phase is shown
      in radians.  Repeatable structure from the Galaxy is visible
      over the three-day time scale in all plots.}
    \label{Fig:albatros2_waterfalls}
  \end{center}
\end{figure}

\subsection{Solar power system}

Although the first-deployed two-element pathfinder is located at a
reasonable walking distance from the main base and uses a manually
operated generator to periodically recharge batteries, the bulk of the
antenna stations that will comprise the full array will be located at
points farther removed from the main base and will therefore require
fully autonomous energy sources.  For the autonomous station
pathfinder, we developed a solar charging system to power the station,
and we are investigating small wind turbines for future
\albatros\ stations.

The autonomous station is powered by two series-connected, 12~V, deep
cycle, lead acid batteries, charged by an array of nine SunPower
SPE-E-Flex-110 solar panels. These panels each have a standard test
capacity of 110~W. Although only $\sim$50~W is required to run the
station, the $\sim$1~kW charging capacity is intentionally oversized
to account for the frequent overcast conditions on Marion.  The
generous power margin allows the charge level to recover quickly on a
short sunny day in winter in the event of charge loss over several
consecutive overcast days.  The nine solar panels are distributed
between three custom designed structures built from aluminum
extrusion, with rigid metalized plastic panels backing the
semi-flexible solar cells. The structures are oriented due north and
are designed to incline the solar panels at a relatively steep angle
to maximize performance under winter conditions, when sunlight hours
are at their minimum. The solar panel mounts have been designed to
withstand frequent gale-force winds on Marion. While the mounting
structures themselves appear to be strong enough, the greatest
challenge in deployment has been to find adequate anchoring in the
volcanic ground.  Each group of three panels is wired in series, and
the three series strings are connected in parallel to a Victron
BlueSolar MPPT 50$\vert$35 charge controller, which optimizes power
transfer from the solar array when charging is required, and also
monitors charge level, reducing output current when the battery bank
is fully charged.

The power logging and control system runs on an Arduino, which logs
information received from the Victron charge controller to an SD card,
and switches power on and off to the readout electronics box. Logged
power data will be used to refine power system requirements for future
autonomous station deployments. The on/off control is necessary to
prevent battery system damage from overly deep discharge. The system
can also be configured to conserve battery power by running on a
schedule between particular hours, typically overnight, when
ionospheric conditions are more favorable.  The power logging and
control system, along with the Victron charge controller and an EMI
filter, are housed together in an aluminum box. The EMI filter reduces
conducted emissions on the photovoltaic side of the solar charge
controller, which could radiate from the solar array and connecting
wires. (Data collected during the night are not at risk of solar
charge controller EMI contamination.)

\begin{figure}[t]
  \begin{center}
    \includegraphics[width=0.7\linewidth]{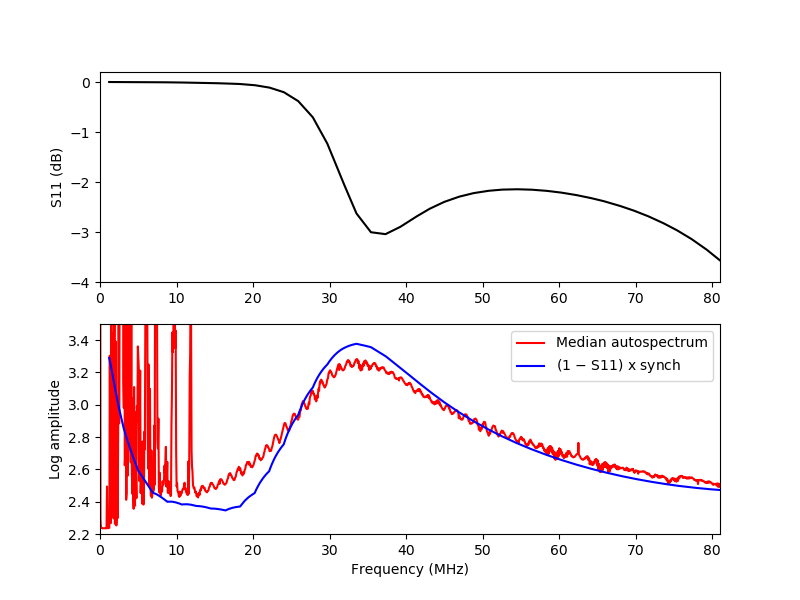}
    \caption{Top panel: simulated $S11$ for the LWA antenna,
      illustrating the steep signal loss below $\sim30$~MHz.  Bottom
      panel: median uncalibrated autospectrum for one of the
      polarizations in the two-element \albatros\ pathfinder, compared
      against a crude sky signal estimate given by the product of $(1
      - S11)$ with a nominal $\nu^{-2.6}$ synchrotron spectrum.  This
      simple model qualitatively demonstrates that the decrease in
      autospectrum power below $\sim30$~MHz is caused primarily by the
      antenna response.}
    \label{Fig:albatros2_auto_model}
  \end{center}
\end{figure}

\begin{figure}[t]
  \begin{center}
    \includegraphics[width=0.8\linewidth]{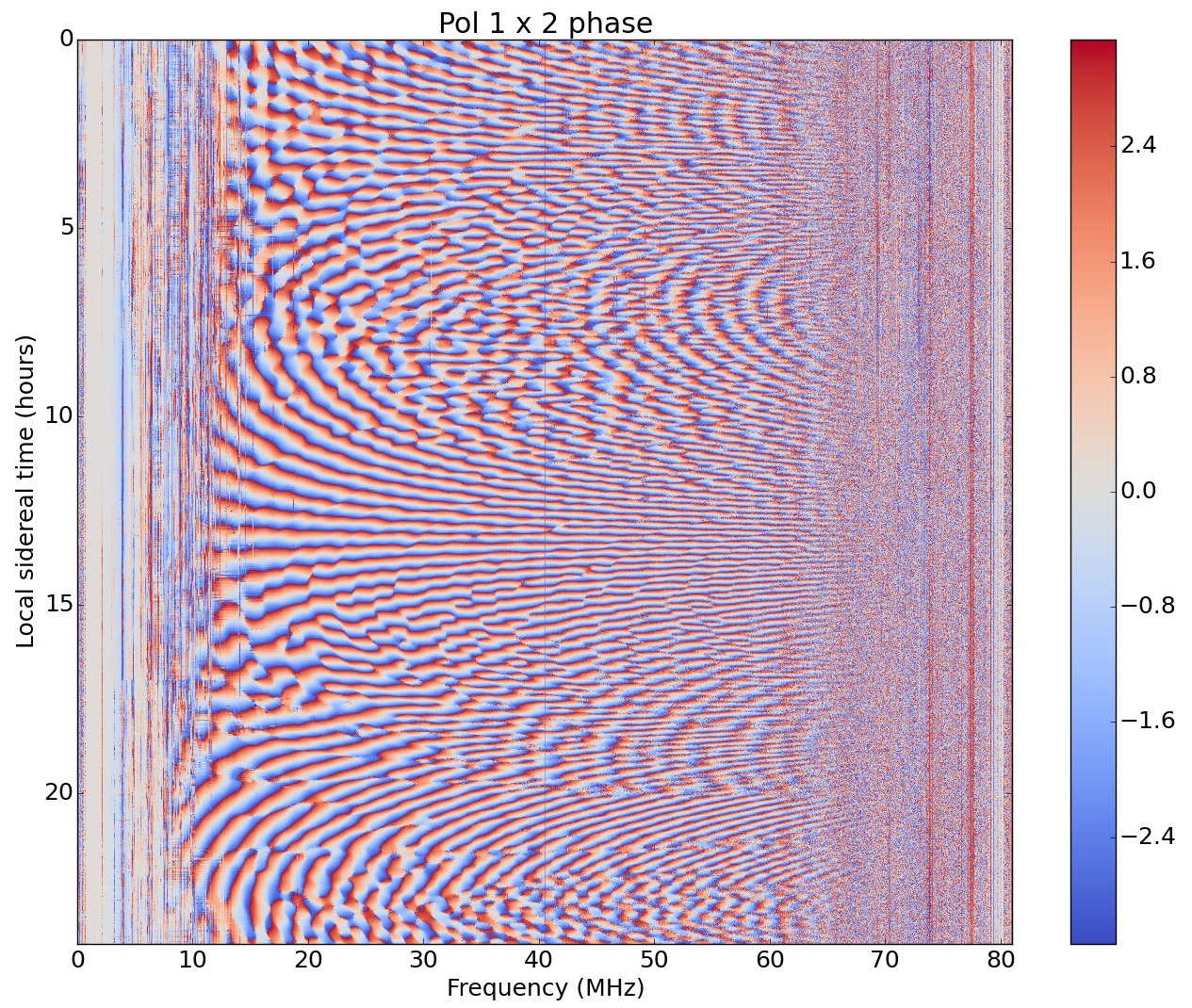}
    \caption{Phases from the cross-correlation between two co-aligned
      polarizations in the two-element \albatros\ pathfinder.  About
      372 hours of data are shown here, binned in local sidereal
      time, and the color stretch is in radians.}
    \label{Fig:albatros2_phase}
  \end{center}
\end{figure}

\section{Preliminary observations}\label{s:obs}

The primary goal of the two-element pathfinder described in
\S\ref{s:2elem} was to qualitatively assess the level of RFI and
ionospheric contamination at low frequencies.
\autoref{Fig:albatros2_waterfalls} illustrates auto- and cross-spectra
from one of the co-aligned polarization pairs from the two antennas.
The data are shown over a three-day period (June 18--21, 2018), and no
calibration or cuts have been applied.  The rising and setting of the
Galaxy is visible in all of the spectrum magnitudes, and repeatable
fringes are visible in the phases of the cross-spectra.  The signal
drop-off below $\lesssim30$~MHz in the autospectra is caused largely
by the combined response of the LWA antenna and FEE, as illustrated in
\autoref{Fig:albatros2_auto_model}.  Although the LWA antenna is not
optimized for the lowest \albatros\ observing frequencies, as
described in \S\ref{s:antenna}, it is sufficient for these initial,
exploratory measurements.  The RFI lines at $\lesssim20$~MHz arise
primarily from shortwave radio transmission reflecting off the
ionosphere and are roughly a factor of two fainter in the
cross-spectrum magnitude in comparison to the autospectrum.  During
the night, when the ionospheric plasma cutoff frequency drops, the
maximum frequency of the RFI contamination falls to $\sim10$~MHz.

We can estimate the instrumental gain stability by comparing the total
power between the different days shown in
\autoref{Fig:albatros2_waterfalls}.  We find that the RMS gain
fluctuations between 30 and 40~MHz (where the Galactic signal peaks in
the data) are less than 1\%.  The typical noise on $\sim$3-second time
scales is roughly 0.04\% of the power.  The low noise suggests that
relative calibration, using autospectra recorded by each autonomous
station, can yield sub-percent accuracy.  (Absolute calibration in the
future will be derived by comparing autospectra against \prizm, which
is co-located and absolutely calibrated, or the Global Sky Model.)  As
a qualitative illustration of the sky signal repeatability on smaller
scales, \autoref{Fig:albatros2_phase} shows the phases of the
cross-spectra binned in local sidereal time.  About 372 hours of data
are averaged into this plot, and no RFI excision has been performed.
A more detailed analysis of signal repeatability will be presented in
a future paper, but the high signal-to-noise fringe pattern, which is
even visible slightly below $\sim10$~MHz, demonstrates the proof of
concept for building the expanded \albatros\ array.

We have been unable to retrieve data from the single autonomous
station pathfinder on Marion (\S\ref{s:autonomous}) because of
COVID-19 restrictions and limited bandwidth to the island, although we
have confirmed that the solar power system has successfully met the
power demands of the antenna station.  To demonstrate baseband writing
capability in the absence of Marion data, we instead present a short
observation that was taken by a separate, single-element
\albatros\ pathfinder station that operated from the McGill Arctic
Research Station (MARS; \ang{79;26;}N, \ang{90;46;}W) on Axel Heiberg
Island, Nunavut, during July~2019.  The RF signal chain of the MARS
installation had the same architecture as that shown in
\autoref{Fig:albatros1_schem}, but the power was supplied by manually
charged batteries, rather than a solar power system.  The MARS RFI
environment is somewhat noisier than that of Marion Island, and
because the data were taken during the height of Arctic summer, the
ionospheric plasma cutoff frequency is higher than the typical cutoff
reflected in the Marion data.  Despite the differences in the power
configuration and RFI environment, the MARS installation is suitable
for illustrating baseband data writing.

\begin{figure}[t]
  \begin{center}
    \includegraphics[width=\linewidth]{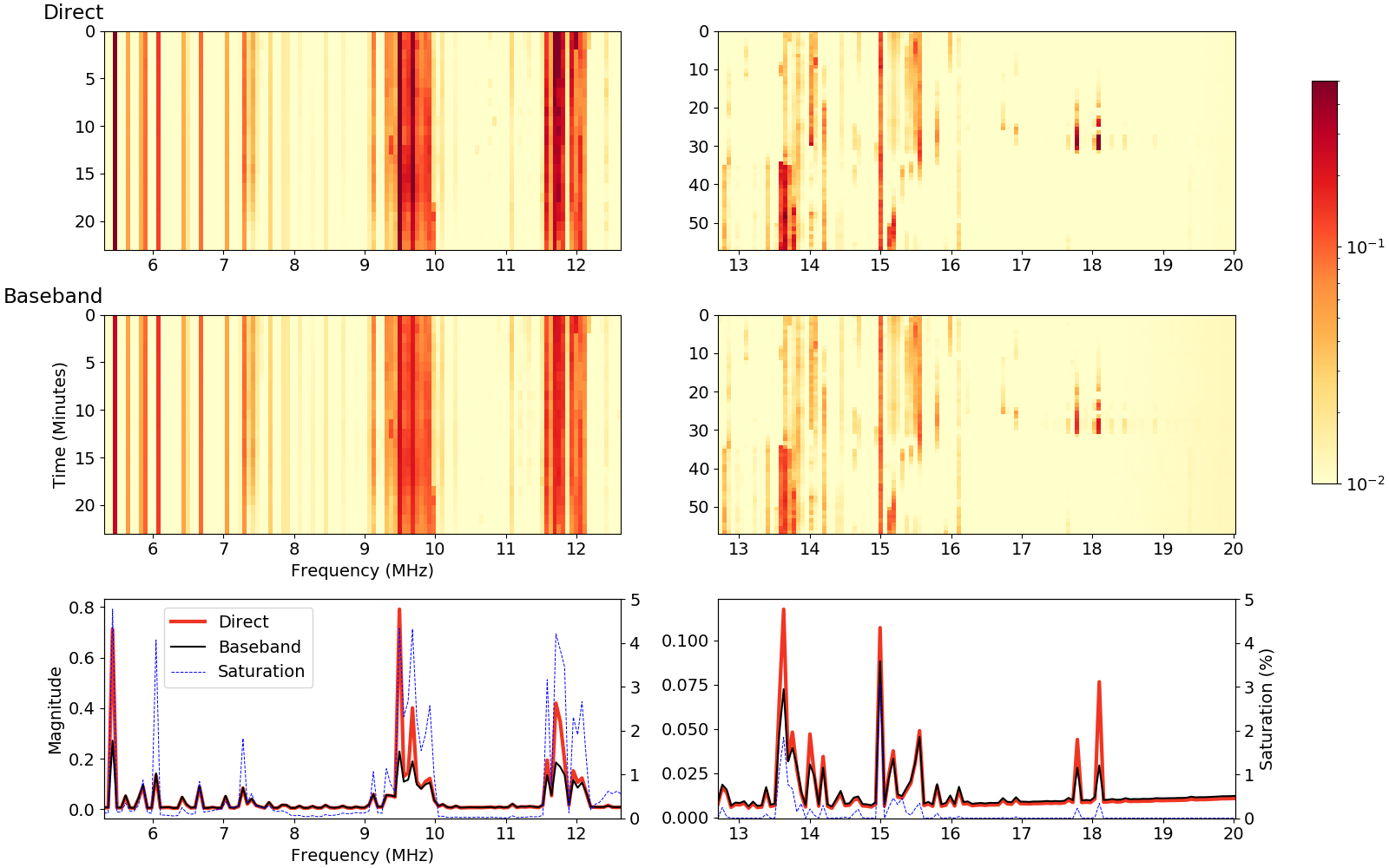}
    \caption{Autospectra for one polarization from an
      \albatros\ pathfinder at MARS (see text for details).  The top
      row shows spectra that are accumulated directly by the SNAP
      board, and the middle row shows spectra computed from baseband
      data with 4-bit quantization.  The two frequency windows,
      5.3--12.6~MHz and 12.6--20.0~MHz, are recorded at two separate
      times, and all plots are shown with 61-kHz resolution.  The
      waterfall plots are normalized so that the maximum is unity and
      are shown with a logarithmic color scale to highlight the
      qualitative agreement in spectral features between the
      directly-accumulated and baseband spectra.  The bottom row
      compares the mean autospectra, which agree well when the
      baseband data are not saturated by bright RFI emission.}
    \label{Fig:baseband_direct_auto}
  \end{center}
\end{figure}

\begin{figure}[t]
  \begin{center}
    \includegraphics[width=0.9\linewidth]{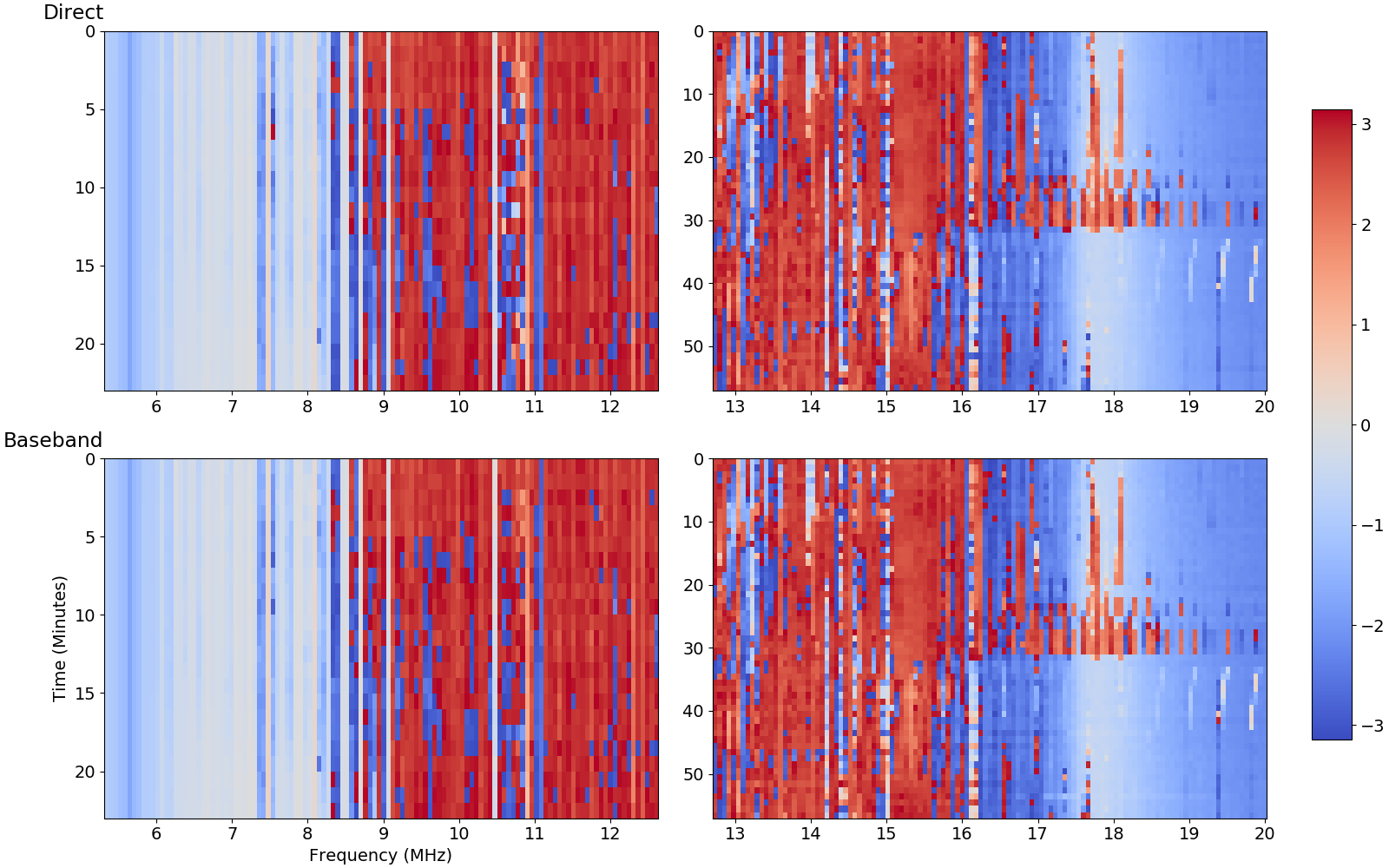}
    \caption{Phases from the cross-correlation between two orthogonal
      polarizations from an \albatros\ pathfinder at MARS (see text
      for details).  The top row shows cross-spectrum phases that are
      accumulated directly by the SNAP board, and the bottom row shows
      phases computed from baseband data with 4-bit quantization.  The
      two frequency windows, 5.3--12.6~MHz and 12.6--20.0~MHz, are
      recorded at two separate times, and all waterfall plots are
      shown with 61-kHz resolution.}
    \label{Fig:baseband_direct_phase}
  \end{center}
\end{figure}

We present short segments of baseband data that were recorded over two
adjacent frequency windows, 5.3--12.6~MHz and 12.6--20.0~MHz, at
separate times.  The baseband data were recorded with 4-bit
quantization to preserve autospectrum information, which is not
available with 1-bit quantization.  \autoref{Fig:baseband_direct_auto}
shows autospectra for one polarization of the \albatros\ pathfinder
installed at MARS, comparing the accumulated spectra that are directly
computed by the SNAP board against spectra that are calculated from
baseband data.  All of the waterfall plots have the same 61-kHz
frequency resolution, set by the resolution of the spectra computed by
the SNAP board.  The waterfall plots are shown with a logarithmic
scale to highlight the qualitative agreement between the
directly-accumulated and baseband spectra for bright and faint
spectral features.  The bottom panels show the time-averaged spectra
on a linear scale to illustrate differences in bright RFI features
that saturate the baseband data.  The fraction of baseband data with
values at the 4-bit extrema are overplotted with spectra, and there is
agreement between the directly-accumulated and baseband spectra except
when significant saturation is present.
\autoref{Fig:baseband_direct_phase} shows waterfall plots of phases
computed from the cross-correlation between the two orthogonal
polarizations of the \albatros\ pathfinder at MARS, comparing the
directly accumulated and baseband data.  Both sets of recorded data
qualitatively agree, up to small differences caused by the 4-bit
quantization.

\section{Summary and future outlook}

We have presented the design of \albatros, a new interferometer that
will image the radio sky at $\lesssim30$~MHz using an array of
autonomous antenna stations installed on Marion Island.  With a
two-element, directly correlated pathfinder, we have demonstrated that
clear, repeatable sky signal is visible from Marion down to
$\lesssim10$~MHz with no data processing or cuts.  We have constructed
the first prototype autonomous \albatros\ station powered by solar
panels, and we have successfully tested the electronics and software
for recording baseband data.

With the proof of concept demonstrated from the pathfinder instruments
presented here, we plan to improve the design of the future
\albatros\ stations that will be installed at the coastal huts on
Marion.  The LWA antenna and FEE are currently not optimized for the
lowest \albatros\ observing frequencies, and we are investigating
possible modifications to the antenna and FEE designs to improve the
low frequency response, as well as building in calibration circuitry.
Because the future \albatros\ stations will be located farther from
the base and will be more difficult to access on a regular basis, each
station will require larger total hard drive volume to store baseband
data over extended periods of time.  We are developing a custom
low-power hard drive bank with $>100$~TB total storage capacity that
will employ a USB multiplexer to select and power only one hard drive
at a time.  The autonomous station pathfinder presented here uses
solar panels to charge the batteries, and we are investigating the use
of small wind turbines as a possible alternative for future stations.
We are also developing analysis tools to compute time-domain data from
the recorded channelized baseband by inverting the polyphase filter
bank while minimizing artifacts from quantization and saturation.

In addition to Marion Island, we are also planning on installing a
second \albatros\ array at MARS in the high Arctic.  As described in
\S\ref{s:obs}, a single pathfinder antenna was installed in July 2019
and observed for about three weeks to assess the RFI environment and
ionospheric conditions.  When fully operational, the Marion and MARS
\albatros\ installations will provide new views of the low-frequency
sky across both hemispheres.

\section*{Acknowledgments}

We gratefully acknowledge the National Research Foundation (grant
number 110929), the South African National Antarctic Programme, the
Natural Sciences and Engineering Research Council of Canada (grant
number RGPNS-2019-534549), and the Polar Continental Shelf Program for
providing funding and logistical support for our research program.
This research was undertaken, in part, thanks to funding from the
Canada 150 Program.  This research was enabled in part by support
provided by SciNet(https://www.scinethpc.ca/) and Compute Canada
(www.computecanada.ca).  The financial assistance of the South
African SKA Project (SKA SA) towards this research is hereby
acknowledged (\url{www.ska.ac.za}).  We additionally extend our
sincere gratitude to the staff at SKA SA for hosting our lab tests.
We thank the South African National Space Agency for their technical
support, the crew of Ultimate Heli for safely delivering us and our
cargo, and all of the Marion takeover and winter team members who
provided invaluable advice and field help.  The authors also wish to
thank members of the LWA and RAPID collaborations for useful
discussions, and Raul Monsalve for helping make our first MARS
campaign a success.

\bibliographystyle{apj}
\bibliography{albatros_paper}{}

\end{document}